\documentclass[aps,pre,twocolumn,superscriptaddress,showpacs]{revtex4}
\usepackage{amssymb}
\usepackage{epsfig}
\usepackage{amsmath}
\usepackage{times}

\begin{document}

\title{Transition to amplitude death in scale-free networks}
\author{Weiqing Liu}
\affiliation{Department of Physics, National University of
Singapore, Singapore 117542} \affiliation{Beijing-Hong
Kong-Singapore Joint Centre for Nonlinear \& Complex Systems
(Singapore), National University of Singapore, Kent Ridge, 119260,
Singapore} \affiliation{Faculty of Science, Jiangxi Univerisity of
Science and Technology, Ganzhou, China, 341000}
\author{Xingang Wang}
\email[Corresponding author. Email address: ]{wangxg@zju.edu.cn}
\affiliation{Institute for Fusion Theory and Simulation, Zhejiang
University, Hangzhou, China 310027}
\author{Shuguang Guan}
\affiliation{Temasek Laboratories, National University of Singapore,
117508, Singapore} \affiliation{Beijing-Hong Kong-Singapore Joint
Centre for Nonlinear \& Complex Systems (Singapore), National
University of Singapore, Kent Ridge, 119260, Singapore}
\author{Choy Heng Lai}
\affiliation{Department of Physics, National University of
Singapore, Singapore 117542} \affiliation{Beijing-Hong
Kong-Singapore Joint Centre for Nonlinear \& Complex Systems
(Singapore), National University of Singapore, Kent Ridge, 119260,
Singapore}

\begin{abstract}
Transition to amplitude death in scale-free networks of nonlinear
oscillators is investigated. As the coupling strength increases, the
network will undergo three stages in approaching to the state of
complete amplitude death. The first stage is featured by a
\emph{``stair-like"} distribution of the node amplitude, and the
transition is accomplished by a \emph{hierarchical death} of the
amplitude stairs. The second and third stages are characterized by,
respectively, a continuing elimination of the synchronous clusters
and a fast death of the non-synchronized nodes.
\end{abstract}

\date{\today }
 \maketitle

The discoveries of the small-world and scale-free properties in
natural and man-made systems have brought a new surge to the study
of collective behaviors in coupled dynamical systems
\cite{SW:1998,BA:1999,CN:REV}, where the scope of the traditional
studies had been significantly extended and a number of new
phenomena had been identified \cite{SYN:REV}. A typical example
could be the synchronization of populations of coupled nonlinear
oscillators. Recent studies have shown that, due to the shorted
network diameter, small-world networks in general have the higher
synchronizability than regular networks of the same parameters
\cite{SYN:SW}. In studying scale-free networks, it has been found
that synchronization is much influenced by the few large-degree
nodes than the majority small-degree ones \cite{SYN:SFN}. Nowadays,
the interplay between the network topology and dynamics had been one
of the focusing issues in nonlinear studies, and there are many
questions waiting for explorations \cite{SYN:REV}.

Amplitude death (AD) refers to the cessation of oscillation of
coupled oscillators when their parameters are considerably
mismatched or there exists time delay in their couplings
\cite{AD:TWO,AD:DELAY-1,AD:DELAY-2}. Typical examples include the
arrhythmia of the cardiac pacemakers \cite{AD:CARDIAC}, the quenched
oscillation of coupled electronic circuits \cite{AD:CIRCUITS},
optical oscillators \cite{AD:OPTICS}, and chemical reactors
\cite{AD:CHEMISTRY}. Recently, this study has been extended to
spatio-temporal systems, where the influences of the frequency
distribution and network topology have been addressed. A chain of
nonlinear oscillators of monotonic frequency distribution has been
studied in Ref. \cite{PAD:LATTICE}, where partial amplitude death
(PAD) (a state where only partial of the network nodes are dead) has
been observed before the complete amplitude death (CAD) of the
system. The authors have also found that, by disordering the
frequency distribution, the generation of CAD can be significantly
postponed. Eliminating CAD by disordered network structure has been
studied in Ref. \cite{AD:SW}, where the small-world networks have
been found to be more resistive to AD than the regular and random
networks. The route to CAD in an array of oscillators has been
investigated in \cite{AD:RING}, where the transition is found to be
divided into different stages.

The papers cited above, however, deal with only the case of
homogeneous networks, i.e., nodes in a network have the similar
degree. As practical systems usually take the form of scale-free
networks characterized by the existence of few very large-degree
nodes, it is intriguing to see how the heterogeneous distribution of
the node degree will affect the generation of AD. In the present
work, by investigating the transition of scale-free networks to CAD,
we shall address the important role of node degree played in AD
generation.

We consider $N$ coupled Landau-Stuart oscillators of the following
form:
\begin{equation}
\dot{Z_{i}}(t)=[r^2+i\omega_{i}-|Z_{i}|^{2}]Z_{i}(t)+\varepsilon\sum_{j=1}^{N}a_{ij}(Z_{j}(t)-Z_{i}(t)),
\label{LS}
\end{equation}
where $i=1,2,\ldots,N$ is the node index. $Z_{i}(t)$ is a complex
number denoting the state of the $i$th oscillator at time $t$,
$\varepsilon$ is the uniform coupling strength, $\omega_{i}$ is the
natural frequency of node $i$, and $r$ is the oscillation growth
rate. The connections of the oscillators are defined by matrix
$A=\{a_{ij}\}$, where $a_{ij}=1$ if $i$ and $j$ are connected, and
$a_{ij}=0$ otherwise. The degree of the $i$th node is
$k_i=\sum_{j=1}^N a_{ij}$. To differentiate the node dynamics,
$\omega_{i}$ is randomly chosen from range $[\omega_1,\omega_2]$.
Without coupling, the trajectory of each oscillator will settle to a
limit circle of radium $|Z_i(t)|=r$. In our study, we will fix all
other parameters in Eq. (\ref{LS}), while increasing $\varepsilon$
to realize the transition.

To measure the degree of the network death, the following two
macroscopic quantities will be employed: The normalized network
``incoherent" energy, $E=\langle \sum_{i=1}^N |Z_i(t)|^2 \rangle /
N$, and the number of the ``dead" nodes $N_d$. Here $\langle \cdot
\rangle$ denotes the time average over time period $T$. It is
important to note that, due to the noise perturbations, in
practice the averaged amplitude of an oscillator, $Z_i=\langle
|Z_i(t)| \rangle$, can not be exactly zero, even through the
condition of AD is fulfilled. So it is necessary to define some
threshold, $Z_0=\langle |Z| \rangle \ll 1$, and regarding nodes of
$Z_i < Z_0$ as dead. Apparently, a state of smaller $E$ and larger
$N_d$ corresponding to a higher degree of amplitude death.

\begin{figure}[tbp]
\begin{center}
\includegraphics[width=0.8\linewidth]{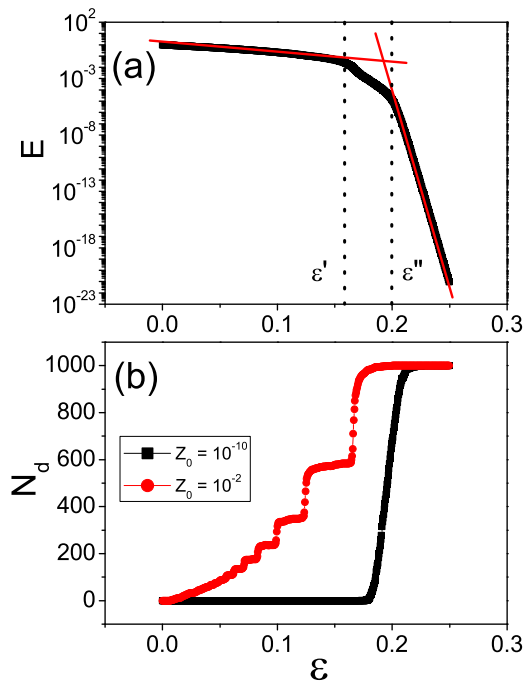}
\caption{(Color online) The transition to CAD for a scale-free
network of $N=1000$ Landau-Stuart oscillators. (a) The variation
of $E$ as a function of $\varepsilon$, where the transition is
roughly divided into three stages by the critical couplings
$\varepsilon' \approx 0.16$ and $\varepsilon'' \approx 0.2$. The
fitted exponents of the first and third stages are $-9\pm 0.1$ and
$-335 \pm 1$, respectively. (b) The variation of $N_d$ as a
function of $\varepsilon$. Two different amplitude thresholds,
$Z_0 = 1 \times 10^{-2}$ (the red curve) and $Z_0 = 1 \times
10^{-10}$ (the black curve), are used separately to characterize
the transition.} \label{Fig_1}
\end{center}
\end{figure}

We first study the transition by means of the macroscopic quantities
$E$ and $N_d$. The network is generated by the standard BA model
\cite{BA:1999}, which consists of $N=1000$ nodes and has average
degree $\langle k \rangle =6$. The degree distribution follows
roughly the power-law scaling $P(k) \sim k^{-3}$. To facilitate the
analysis, we reorder the network nodes by an ascending order of
their degrees. Thus $k_{1}=k_{min}$ and $k_{N}=k_{max}$. For the
local dynamics, we set $r=0.5$ and randomly choose $\omega_i$ from
the range $[1,50]$. To simulate, we initialize the network with
random conditions and evolve it according to Eq. (\ref{LS}). After a
transient period $T'=1 \times 10^3$, we start to calculate $E$ and
$N_d$, which are averaged over another period $T=500$. The variation
of $E$ as a function of $\varepsilon$ is plotted in Fig. 1(a). It is
seen that the evolution of $E$ can be roughly divided into three
stages. In the range $\varepsilon \in [0, 0.16]$, $E$ is decreased
exponentially, but with a relatively slow speed; in the range
$\varepsilon \in (0.16, 0.2)$, the decrease of $E$ is fasted, but
the dependence of $E$ on $\varepsilon$ is complicated (as a matter
of fact, the value of $E$ is sensitively dependent on the frequency
distribution of the oscillators); in the range $\varepsilon \in
[0.2,0.25]$, $E$ is decreased exponentially again, but with a much
faster speed in comparison with the previous stages. The variation
of $N_d$ as a function of $\varepsilon$ is plotted in Fig. 1(b). For
a small amplitude threshold, e.g. $Z_0=1\times 10^{-10}$, $N_d$ is
found to be smoothly increased from 0 to $N$ within a narrow range
of $\varepsilon$. However, if the amplitude threshold is not too
small, e.g. $Z_0>1\times 10^{-2}$, a non-smooth stair-like evolution
of $N_d$ will be presented.

\begin{figure}[tbp]
\begin{center}
\includegraphics[width=0.8\linewidth]{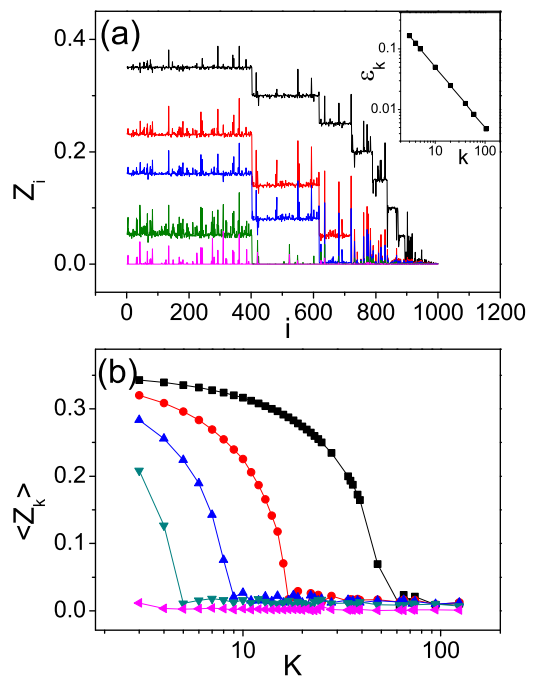}
\caption{(Color online) For the same network used in Fig. 1, typical
amplitude distributions observed in the first stage of the
transition. (a) The distributions are generated by coupling
strengths, from top to bottom, $\varepsilon
=0.05,0.09,0.1,0.15,0.16$. Inset: The inverse relationship between
the critical coupling $\varepsilon_k$ and the node degree $k$. The
symbols are the numerical results, and the solid line is drawn by
the theory. (b) For the same set of couplings as in (a), the
averaged stair amplitude $\langle Z_k \rangle$ versus the node
degree $k$. Each data of (b) is averaged over 50 network
realizations. } \label{Fig_2}
\end{center}
\end{figure}

To explore the staged transition and the stair structures appeared
in Fig. 1, we proceed to investigate the system dynamics at the
microscopic level. Specifically, we shall calculate the
distribution of the node amplitude, $\{Z_i\}$, and see how this
distribution is evolved during the transition. In the first stage,
the node amplitude is evolved as follows. At very small coupling
$\varepsilon\approx0$, all nodes own the same amplitude $Z_i=r$,
regardless of the difference of the node degree. Then, as
$\varepsilon$ increases, all the amplitude will be decreased, but
with very different speed. Specifically, \emph{the decrease of the
amplitude of a node is proportional to the node degree}. So the
amplitude distribution is gradually curved and replaced by a
non-smooth ``stair-like" distribution, say, for example, the one
generated by $\varepsilon=0.05$ in Fig. 2(a). In this stair-like
distribution, nodes of the same degree will present the same
amplitude, except some rare ``bursts" which are caused by
synchronous clusters (to be explained later). Because of the
heterogenous degree distribution, the stairs are of different
length. More interestingly, the height of the stairs is gradually
stepping down with the node degree. That is, the highest stair
consists of only the smallest-degree nodes and the lowest stair
consists of only the largest-degree nodes. Increasing
$\varepsilon$ further, the amplitude of the largest-degree nodes
is hardly decreased, while the amplitude of the other nodes will
be decreased continuously. This results a hierarchical death of
the amplitude stairs. The process of stair elimination is shown in
Fig. 2(a), where some typical amplitude distributions observed in
the first stage are presented. Finally, at about the coupling
strength $\varepsilon'$, the last stair, i.e. the stair consisting
of the smallest-degree nodes, will be eliminated, and the
first-stage transition is completed. For instance, the
distribution generated by $\varepsilon=0.16$ in Fig. 2(a). So the
stair structures appeared in Fig. 1(b) is just a reflection of the
hierarchical death of the amplitude stairs.

The formation of amplitude stairs could be analyzed by the
mean-field approximation. Noticing that Eq. (\ref{LS}) can be
rewritten as $\dot{Z_{i}}(t)=[(r^2-\varepsilon
k_i)+i\omega_{i}-|Z_{i}|^{2}]Z_{i}(t)+G_i$, where
$G_i=\varepsilon\sum_{j=1}^{N}a_{ij}Z_{j}$ is the collective
coupling received by node $i$. When $\varepsilon$ is small, most
oscillators of the network will behave incoherently, therefore $G_i$
is small and negligible to the node dynamics, especially for the
large-degree nodes. With this concern, the node dynamics can be
simplified to $\dot{Z_{i}}(t)=[(r^2-\varepsilon
k_i)+i\omega_{i}-|Z_{i}|^{2}]Z_{i}(t)$. By requiring
$|\dot{Z_{i}}(t)|=0$, we obtain
\begin{equation}
|Z_i|^2=r^2-\varepsilon k_i. \label{STAIR}
\end{equation}
Eq. (\ref{STAIR}) tells that, for a given coupling strength, the
amplitude of a node is only dependent on the node degree. This
explains why the height of the stairs in Fig. 2(a) is decreased with
the node degree. From Eq. (\ref{STAIR}) we can also estimate the
averaged amplitude of each stair $Z_k$, which is verified by the
numerical data of Fig. 2(a). Moreover, by setting $|Z_i|=0$ in Eq.
(\ref{STAIR}), we can obtain the critical coupling $\varepsilon_k$
where the $k$-degree stair is eliminated: $\varepsilon_k \approx
r^2/k $. The inverse relationship between $Z_k$ and $k$ is verified
in the inset plot of Fig. 2(a).

It should be mentioned that the above analysis works for only the
case of weakly coupled networks. If the coupling strength is not too
weak, some nodes in the network could behave coherently. In such a
case, the value of $G_i$ will be deviated from $0$, and the flat
stairs predicted by Eq. (\ref{STAIR}) will be disturbed. This is
just we have observed in simulations. In Fig. 2(a), despite the
changes of the coupling strength, there always exist some bursts in
the distributions. These bursts, which have random locations and
various amplitude, are directly resulted from the synchronized
nodes. The picture is the following. Since the node frequency is
randomly chosen, there could be the situation that some connected
nodes in the network have very small frequency mismatch. As the
coupling strength increases, these nodes will be easily synchronized
and form some synchronous clusters (phase synchronization
\cite{PS}). Once synchronized, nodes will be efficiently protected
from AD \cite{PAD:LATTICE,AD:RING}, showing as the amplitude bursts.
The synchronized nodes, however, may have different amplitude, as
they could be embraced by different set of neighbors. To smooth the
amplitude bursts caused by synchronization, we have calculated the
averaged stair amplitude, $\langle Z_k \rangle$, as a function of
$k$. Now a smooth, hierarchical transition of AD is presented [Fig.
2(b)].

\begin{figure}[tbp]
\begin{center}
\includegraphics[width=\linewidth]{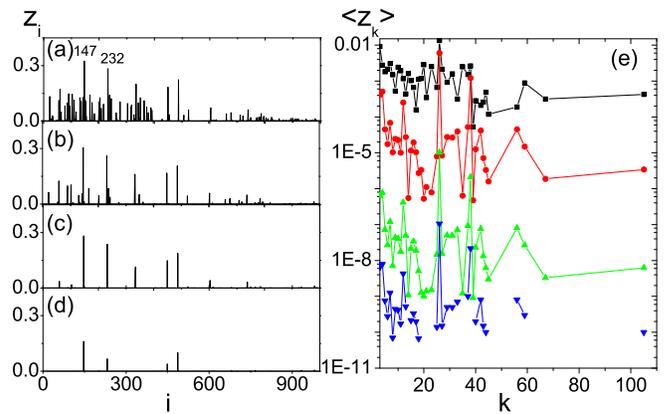}
\caption{(Color online) For a slightly modified network model, the
amplitude distributions observed in the second and third stages of
the transition. (a) $\varepsilon=0.18$ and (b) $\varepsilon=0.19$
are within the second stage. (c) $\varepsilon=0.21$ and (d)
$\varepsilon=0.22$ are within the third stage. (e) For the above
amplitude distributions, the variations of $\langle Z_k \rangle$ as
a function of $k$. The missed data in the last distribution is
caused by the limited computing precision.} \label{Fig_3}
\end{center}
\end{figure}

We go on to investigate the transition in the second stage. At the
end of the first stage, a number of synchronous clusters are formed
in the network. Because of synchronization, the amplitude of the
synchronized nodes are apparently larger than the non-synchronized
ones. In the second stage, as the coupling strength increases,
\emph{both the number and the size of the synchronous clusters are
decreased}. Interestingly, it is found that the robustness of a
cluster is mainly determined by the frequency mismatch between the
synchronized nodes, while is less affected by the node degree or
cluster size. To show this point more clearly, we have slightly
modified the network by artificially adjusting the frequency
mismatch between two connected nodes, $(147,232)$, to be $\delta
\omega = 1\times 10^{-3}$. This pair of nodes, both having the
smallest degree $k_{min}=3$, are synchronized at a very small
coupling and are kept at large amplitude till the very end of the
transition. As shown in Fig. 3, despite of the increase of the
coupling strength, the amplitude of these two nodes is always
apparently larger than the resting nodes. Different to the situation
of the first stage, in the second stage the node amplitude is less
dependent of the node degree, as indicated by Fig. 3(e).

The second-stage transition is ended up with a state of very few
synchronous clusters. Typically, each cluster consists of only
several nodes which have very close natural frequencies, like the
pair of nodes discussed above. The few survival clusters, however,
are extremely robust and could stand for a very large coupling
strength. The robust clusters result the new form of transition in
the third stage: \emph{A fast death of the non-synchronized nodes
accompanied by a slow death of the synchronized nodes}. This
property of network transition can be read from Fig. 3(b) and (c),
where the amplitude of nodes $147$ and $232$ are kept at large
values, while the amplitude of the other nodes is extremely small
(see also Fig. 3(e)). Like the second stage, in the third stage the
node amplitude is also less dependent on the node degree, as can be
found from Fig. 3(e). With the elimination of the most robust
cluster, the whole transition is completed.

\begin{figure}[tbp]
\begin{center}
\includegraphics[width=\linewidth]{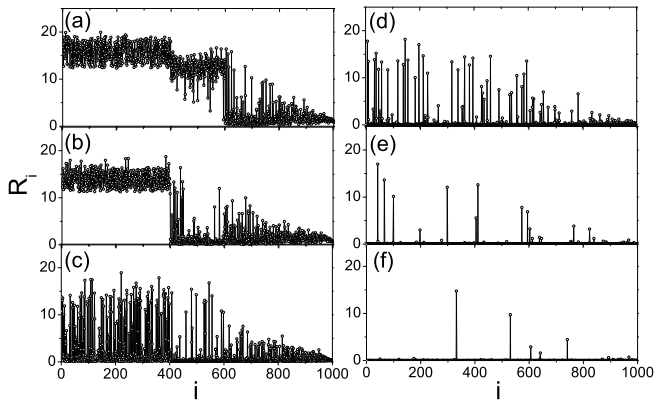}
\caption{Typical amplitude distributions ($R_i=\langle
\sqrt{x_i^2(t)+y_i^2(t)} \rangle$) observed in the transition of the
network of chaotic R\"{o}ssler oscillators. (a) $\varepsilon =
0.039$ and (b) $\varepsilon = 0.05$ are within the first stage; (c)
$\varepsilon = 0.055$ and (d) $\varepsilon = 0.06$ are within the
second stage; and (e) $\varepsilon = 0.065$ and (f) $\varepsilon =
0.07$ are within the third stage. } \label{Fig_4}
\end{center}
\end{figure}

Whether the above phenomena about AD transition can be found for the
general dynamics, say, for example, chaotic oscillators? To answer
this, we have checked the transition of the same network but for
chaotic R\"{o}ssler oscillators. The dynamics of a single oscillator
is described by $F_i(\textbf{x})=[-\omega_i y_i-z_i,\omega_i x_i
+0.165y_i,z_i(x_i-10)+0.2]$, where $\omega_i$ is the natural
frequency of the $i$th oscillator. In simulation, $\omega_i$ is
chosen randomly from the range $[1,3]$. The coupling function is
$H(\textbf{x})=x$. Fig. 4 is a collection of the amplitude
distributions observed during the transition, which repeat the
phenomena of Landau-Stuart oscillators. Besides the specific network
in Fig. 1, we have also tested the transition to CAD in other
networks, including changing the network size and the degree
distribution, adopting the clustered networks, and considering the
degree assortativity. The general finding is that, given the degree
distribution is heterogeneous, the phenomena of staged transition
and stair structures will be presented.

To summarize, we have studied the transition to CAD in scale-free
networks of nonlinear oscillators, and found the important role of
node degree played in AD generation. Since many practical systems
where AD is importantly concerned possess heterogeneous degree
distribution, our findings might give some new thoughts to the
relevant studies, say, for example, the stability and evolution of
ecological networks \cite{MRM:2001,ECOSYS}.

XGW is supported by the National Natural Science Foundation of China
under Grant No. 10805038.


\begin{thebibliography}{99}

\bibitem{SW:1998} D.J. Watts and S.H. Strogatz, Nature \textbf{393},
440 (1998).

\bibitem{BA:1999} A.-L. Barab\'{a}si and R. Albert, Science
\textbf{286}, 509 (1999).

\bibitem{CN:REV} R. Albert and A.-L. Barab\'{a}si, Rev. Mod. Phys.
\textbf{74}, 47 (2002); M.E. Newman, SIAM Rev. \textbf{45}, 167
(2003).

\bibitem{SYN:REV} S. Boccaletti, V. Latora, Y. Moreno, M. Chavez,
and D.-U. Hwang, Phys. Rep. \textbf{424}, 175 (2006); S. Boccaletti
and L.M. Pecora, Chaos \textbf{16}, 015101 (2006); J.A.K. Suykens
and G.V. Osipov, Chaos \textbf{18}, 037101 (2008).

\bibitem{SYN:SW} X. F. Wang and G. Chen, Int. J. Bifurcation Chaos Appl. Sci.
Eng. \textbf{12}, 187 (2002); M. Barahona and L. M. Pecora, Phys.
Rev. Lett. \textbf{89}, 054101 (2002).

\bibitem{SYN:SFN} A. E. Motter, C. S. Zhou, and J. Kurths, Europhys. Lett. \textbf{69},
334 (2005); X.G. Wang, Y.-C. Lai, and C.H. Lai, Phys. Rev. E
\textbf{75}, 056205 (2007).

\bibitem{AD:TWO} Y. Yamaguchi and H. Shimizu, Physica (Amsterdam) \textbf{11D}, 212 (1984);
K. Bar-Eli, Physica D \textbf{14}, 242 (1985); R.E. Mirollo and S.H.
Strogatz, J. Stat. Phys. \textbf{60}, 245 (1989); D.G. Aronson, G.B.
Ermentrout, and N. Kopell, Physica (Amsterdam) \textbf{41D}, 403
(1990); G.B. Ermentrout, Physica (Amsterdam) \textbf{41D}, 219,
(1990).

\bibitem{AD:DELAY-1} D.V. Ramana Reddy, A. Sen, and G.L. Johnston,
Phys. Rev. Lett. \textbf{80}, 5109 (1998); Physica (Amsterdam)
\textbf{129D}, 15 (1999); Phys. Rev. Lett. \textbf{85}, 3381 (2000).

\bibitem{AD:DELAY-2}F.M. Atay, Physica (Amsterdam) \textbf{183D}, 1 (2003); Phys. Rev.
Lett. \textbf{91}, 094101, (2003).

\bibitem{AD:CARDIAC} S.H. Strogatz, Nature (London) \textbf{394},
316 (1998).

\bibitem{AD:CIRCUITS} R. Herrero, M. Figueras, J. Rius, F. Pi, and G.
Orriols, Phys. Rev. Lett. \textbf{84}, 5312 (2000); D.V. Ramana
Reddy, A. Sen, and G.L. Johnston, Phys. Rev. Lett. \textbf{85}, 3381
(2000).

\bibitem{AD:OPTICS} D.V. Ramana Reddy, A. Sen, and G.L. Johnston,
Phys. Rev. Lett. \textbf{80}, 5109 (1998); M. Wei and J. Lun, Appl.
Phys. Lett. \textbf{91}, 061121 (2007).

\bibitem{AD:CHEMISTRY} M.F. Crowley and I.R. Epstein, J.
Phys. Chem. \textbf{93}, 2496 (1989); M. Yoshimoto, K. Yoshikawa,
and Y. Mori, Phys. Rev. E \textbf{47}, 864 (1993).

\bibitem{PAD:LATTICE} L. Rubchinsky and M. Sushchik, Phys. Rev. E
\textbf{62}, 6440 (2000).

\bibitem{AD:RING} J. Yang, Phys. Rev. E \textbf{76}, 016204 (2007).

\bibitem{AD:SW} Z. Hou and H. Xin, Phys. Rev. E \textbf{68},
055103(R) (2003).

\bibitem{PS} M.G. Rosenblum, A.S. Pikovsky, and Jurgen Kurths, Phys.
Rev. Lett. \textbf{76}, 1804 (1996).

\bibitem{MRM:2001} R.M. May, (2001) \emph{Stability and Complexity in Model
Ecosystems.} Princeton Landmarks in Biology. Princeton: Princeton
University Press, (2001).

\bibitem{ECOSYS}J.M. Montoya and R. V. Sol\'{e}, preprint
cond-mat/0011195; Camacho, J.R. Guimer\`{a} , and L.A.N. Amaral,
preprint cond-mat/0102127.

\end{thebibliography}
\end{document}